\documentclass[runningheads]{svmult}

\usepackage{makeidx}   
\usepackage{graphicx}  
\usepackage{subeqnar}  
\usepackage{multicol}  
\usepackage{physprbb}  
\usepackage{amsmath,amssymb}
\makeindex             

\usepackage{latexsym}

\begin{document}
\title*{Gravitomagnetism and the Clock Effect}
\toctitle{Gravitomagnetism and the Clock Effect}
%
%
\titlerunning{}
%
\author{Bahram Mashhoon\inst{1}
\and Frank Gronwald \inst{2}
\and Herbert I.M.\ Lichtenegger \inst{3}}
\authorrunning{Bahram Mashhoon et al.}
%
%
\institute{Department of Physics and Astronomy, University of
Missouri, Columbia, Missouri, 65211, USA
\and Institut GET, 
Universit\"{a}t Magdeburg, D-39106 
Magdeburg, Germany
\and Institut f\"{u}r Weltraumforschung, \"{O}sterreichische Akademie der
Wissenschaften, A-8010 Graz, Austria}

\maketitle              

\begin{abstract}
The main theoretical aspects of gravitomagnetism are reviewed.  It is shown
that the gravitomagnetic precession
of a gyroscope is intimately connected with the special temporal structure
around a rotating mass that is revealed
by the gravitomagnetic clock effect.  This remarkable effect, which
involves the difference in the proper periods
of a standard clock in prograde and retrograde circular geodesic orbits
around a rotating mass, is discussed in detail.
The implications of this effect for the notion of ``inertial dragging'' in
the general theory of relativity are presented.
The theory of the clock effect is developed within the PPN framework and
the possibility of measuring it via spaceborne
clocks is examined.
\end{abstract}

\section{Introduction}

The close formal similarity between Coulomb's law of electricity and
Newton's law of gravitation has led to a description
of Newtonian gravitation in terms of a gravitoelectric field.  The
classical tests of general relativity can all be described
via post-Newtonian gravitoelectric corrections brought about by relativity
theory.  Moreover, any theory that combines
Newtonian gravitation and Lorentz invariance in a consistent framework must
involve a gravitomagnetic field in close analogy
with electrodynamics.  The gravitomagnetic field is generated by the motion
of matter.  For instance, the mass current in the
rotating Earth generates a dipolar gravitomagnetic field that has not yet
been directly observed; in fact, the main objective
of the GP-B is to measure this field in a polar Earth orbit via the
gravitomagnetic precession of superconducting gyroscopes on
board a drag-free satellite.

Gravitomagnetism had its beginning in the second half of the last century.
Developments in electrodynamics led Holzm\"{u}ller
\cite{GHo} and Tisserand \cite{FTi} to postulate the existence of a solar
gravitomagnetic field \cite{ande94}. In fact, attempts were made to account
for the excess perihelion precession of Mercury since the planetary orbits
would be affected by the gravitomagnetic field
of the Sun.  However, the excess perihelion precession of Mercury was
successfully explained by Einstein's general relativity
theory in terms of a small relativistic correction to the Newtonian
gravitoelectric potential of the Sun.  It was later shown
by Thirring and Lense \cite{HTh,BM1} that general relativity also predicts
a certain gravitomagnetic field for a rotating mass, but the
magnitude of this field in the solar system is generally small and would
lead to a retrograde precession of the planetary
orbits.  This Lense-Thirring precession of planetary orbits is too small to
be detectable at present.

For the purposes of confronting the theory with observation,
gravitomagnetic phenomena are usually described in the framework
of the post-Newtonian approximation; however, it is possible to provide a
fully covariant treatment of certain
aspects of gravitoelectromagnetism
\cite{AMa,BJH}.  In fact, extensions of the Jacobi equation (i.e. the
relativistic tidal equation) may be employed to identify the
gravitoelectric and gravitomagnetic components of the curvature tensor in
close analogy with the Lorentz force law.  This
analogy is incomplete, however, since the purely spatial components of the
curvature tensor do not in general have an analog
in the electromagnetic case; in fact, this is expected since linear
gravity is a spin-2 field in 
contrast to the spin-1 character of the electromagnetic field.

Some of the main theoretical aspects of gravitomagnetism are discussed
in Section 2.  We then turn our attention to how gravitomagnetism
affects the spacetime structure in general relativity.  Of primary
importance in this connection is the gravitomagnetic clock effect,
which in its simplest form may be formulated in terms of the
difference in the proper periods of two clocks moving on the same
circular orbit but in opposite directions about a rotating mass.  Let
$\tau_{+}(\tau_{-})$ be the period for prograde (retrograde) motion,
then for $r\gg2GM/c^{2},\tau_{+} - \tau_{-} \approx 4\pi J/(Mc^{2})$.
To lowest order, this remarkable result is independent of Newton's
constant of gravitation $G$ and the radius of the orbit $r$.  The
effect and its consequences are discussed in Section 3 for circular
equatorial orbits in the Kerr geometry and the intimate connection
between the clock effect and the gravitomagnetic gyroscope precession
is demonstrated.  The PPN approximation for this effect is developed
in Section 4 and a brief discussion of its observability is given in
Section 5.  The sign of the clock effect is quite intriguing, as it
implies that prograde equatorial clocks are slower than retrograde
equatorial clocks.  This is completely opposite to what would be
expected on the basis of ``inertial dragging''.  In fact,
gravitomagnetism is historically connected with the question of the
origin of inertia as this was Thirring's motivation in his original
paper on gravitomagnetism \cite{HTh}.  The present status of the problem of
inertia is the subject of Section 6.  Finally, Section 7 contains a
brief discussion.

Unless specified otherwise, we use units such that $G=c=1$ for the sake of
convenience.

\section{Gravitoelectromagnetism}

This section is devoted to a brief discussion of certain essential theoretical
aspects of gravitoelectromagnetism.  The Larmor theorem has played an
important role in the field of magnetism; therefore, we begin by an account of
the gravitational analog of Larmor's theorem.

\vspace{.15in}

\noindent {\it Gravitational Larmor Theorem}

A century ago, Larmor established a theorem regarding the local equivalence of
magnetism and rotation \cite{1}.  That is, the basic electromagnetic force on a
slowly moving particle of charge $q$ and mass $m$ can be locally replaced in
the linear approximation by the inertial forces that arise if the motion is
referred instead to an accelerated system in the absence of the electromagnetic
field.  The translational acceleration of the system is related to the electric
field, ${\bf a}_{L}=-(q/m) {\bf E}$, and the rotational (Larmor) frequency is
related to the magnetic field via $\mbox{\boldmath $\omega$}_{L}=q{\bf B}/(2mc)$. 
The charge-to-mass ratio is not the same for all particles; otherwise, a geometric
theory of electrodynamics could be developed along the same lines as general
relativity.  It turns out that in general relativity one can provide an
interpretation of Einstein's heuristic principle of equivalence via the
gravitational Larmor theorem \cite{2}.  This is due to the experimentally
well-established circumstance that the gravitational charge-to-mass ratio is the
same for all particles.  Einstein's heuristic principle of equivalence is usually
stated in terms of the gravitoelectric field, i.e. the translational acceleration
of the ``Einstein elevator'' in Minkowski spacetime.  The gravitational Larmor
theorem would also involve the gravitomagnetic field, i.e. a rotation of the
elevator as well.

It follows from the theoretical study of the motion of test particles as well
as ideal test gyroscopes in a gravitational field that in general relativity
the gravitoelectric charge is $q_{E}=-m$, while the gravitomagnetic charge is
$q_{B}=-2m$; in fact, $q_{B}/q_{E}=2$ since general relativity involves the
tensor potential $g_{\mu \nu}$, i.e. (linear) gravitation is a spin-2 field. 
Thus ${\bf a}_{L}={\bf E}$ and $\mbox{\boldmath $\omega$}_{L}=-{\bf B}/c$ in
this case.  Indeed ${\bf B}/c= \mbox{\boldmath $\Omega$}_{P}$ is the gravitomagnetic
precession frequency of an ideal test gyroscope at rest in a gravitomagnetic field,
i.e. far from a rotating source $d{\bf S}/dt=\mbox{\boldmath $\Omega$}_{P} \times
{\bf S}$, where

\begin{equation}
\mbox{\boldmath $\Omega$}_{P} = \frac{GJ}{c^{2}r^{3}} [3 (\hat{\bf r} \cdot
\hat{\bf J} ) \hat{\bf r} - \hat{\bf J}],
\end{equation}

\noindent and $J$ is the total angular momentum of the source.  Let us note that a
gyro spin is in effect a gravitomagnetic dipole moment that precesses in a
gravitomagnetic field.  Locally, the same rotation would be observed in the
absence of the gravitomagnetic field but in a frame rotating with frequency
$\mbox{\boldmath $\omega$}_{L}=-\mbox{\boldmath $\Omega$}_{P}$ in agreement with the
gravitational Larmor theorem.

It is the goal of the GP-B to measure the gravitomagnetic gyroscope precession
in a polar orbit about the Earth and thereby provide {\it direct}
observational proof of the existence of the gravitomagnetic field \cite{3}.

\vspace{.15in}

\noindent {\it Gravitoelectromagnetic Field}

Let us consider the gravitational field of a ``nonrelativistic'' rotating
astronomical source in the linear approximation of general relativity.  The
spacetime metric may be expressed as $g_{\mu \nu}=\eta_{\mu \nu}+h_{\mu \nu}$,
where $\eta_{\mu \nu}$ is the Minkowski metric.  We define $\bar{h}_{\mu
\nu}=h_{\mu \nu} - {1 \over 2} \eta_{\mu \nu} h$, where 
$h={\rm tr}(h_{\mu \nu})$; then,
the gravitational field equations are given by

\begin{equation}
\Box \bar{h}_{\mu \nu} = -16 \frac{\pi G}{c^{4}} T_{\mu \nu},
\label{feq}
\end{equation}

\noindent where the Lorentz gauge condition $\bar{h}^{\mu \nu},_{\nu} = 0$ has
been imposed.  We focus attention on the particular retarded solution of the field
equations given by

\begin{equation}
\bar{h}_{\mu \nu} = \frac{4G}{c^{4}} \int \frac{T_{\mu \nu} (ct-| {\bf x} - {\bf
x}^{\prime} |, \; {\bf x}^{\prime})}{| {\bf x} - {\bf x}^{\prime} |}\: d^{3}
x^{\prime},
\end{equation}

\noindent where the nature and distribution of the ``nonrelativistic'' source must
be taken into account.

We are interested in sources such that $\bar{h}_{00}=4 \Phi / c^{2}$,
$\bar{h}_{0i}=-2A_{i}/c^{2}$ and $\bar{h}_{ij}=O(c^{-4})$, where $\Phi(t,{\bf x})$
is the gravitoelectric potential, ${\bf A}(t,{\bf x})$ is the gravitomagnetic
vector potential and we neglect all terms of order $c^{-4}$ and lower including
the tensor potential $\bar{h}_{ij}(t,{\bf x})$.  It follows that
$T^{00}/c^{2}=\rho$ is the effective gravitational charge density and
$T^{0i}/c=j^{i}$ is the corresponding current.  Thus, far from the source

\begin{equation}
\Phi \sim \frac{GM}{r}, \;\;\;\;\;\;\;\;\;\;\; {\bf A} \sim \frac{G}{c} \frac{{\bf
J} \times {\bf r}}{r^{3}},
\end{equation}

\noindent where $M$ and ${\bf J}$ are the total mass and angular momentum of the
source, respectively.  It follows from the Lorentz gauge condition that

\begin{equation}
\frac{1}{c} \frac{\partial \Phi}{\partial t} + \mbox{\boldmath $\nabla$} \cdot
\left( {1\over 2} {\bf A} \right) = 0,
\label{gauge}
\end{equation}

\noindent since the other three equations $( \bar{h}^{i\mu},_{\mu} = 0)$ all
involve terms that are of $O(c^{-4})$ and therefore neglected.  The spacetime
metric involving the gravitoelectromagnetic (``GEM'') potentials is then 
given by

\begin{equation}
-ds^{2}=-c^{2} \left( 1- {2\over c^{2}} \Phi \right) dt^{2} - {4\over c} ({\bf A}
\cdot d{\bf x} ) dt + \left( 1+ {2\over c^{2}} \Phi \right) \delta_{ij}dx^{i}
dx^{j}. \label{e6}
\end{equation}

The GEM fields are defined by

\begin{equation}
{\bf E} =- \mbox{\boldmath $\nabla$} \Phi - {1 \over c} {\partial \over \partial
t} \left( {1\over 2} {\bf A} \right), \;\;\;\;\;\;\;\; {\bf B} = \mbox{\boldmath
$\nabla$}
\times {\bf A}, \label{gem}
\end{equation}

\noindent in close analogy with electrodynamics.  It follows from the field
equations (\ref{feq}) and the gauge condition (\ref{gauge}) that

\begin{equation}
\mbox{\boldmath $\nabla$} \cdot {\bf E} = 4 \pi G \rho,
\label{e1}
\end{equation}

\begin{equation}
\mbox{\boldmath $\nabla$} \cdot \left( {1\over 2} {\bf B} \right) = 0,
\label{e2}
\end{equation}

\begin{equation}
\mbox{\boldmath $\nabla$} \times {\bf E} = - {1\over c} {\partial \over \partial t}
\left( {1 \over 2} {\bf B} \right),
\label{e3}
\end{equation}

\begin{equation}
\mbox{\boldmath $\nabla$} \times \left( {1\over 2} {\bf B} \right) = {1\over
c}{\partial \over \partial t} {\bf E} + {4 \pi \over c} G \:{\bf j}\:,
\label{e4}
\end{equation}

\noindent which are the Maxwell equations for the GEM field.  Using classical
electrodynamics as a guide, one can investigate the various
implications of these equations \cite{4}.  A thorough approach to the
determination of the gravitomagnetic field of a rotating mass (such as
the Earth) is contained in the papers of Teyssandier \cite{5}.

The fact that the magnetic parts of equations (\ref{e1}) - (\ref{e4}) always
appear with a factor of $1/2$ as compared to standard electrodynamics
is due to the circumstance that the effective gravitomagnetic charge
is twice the gravitoelectric charge.  That is, $Q_{E} = M$ and
$Q_{B}=2M$ are the effective gravitoelectric and gravitomagnetic
charges of the source.

The linear approximation of general relativity involves a spin-2
field.  This field, once its spatial components are neglected, can be
interpreted in terms of a gravitoelectromagnetic vector potential.  To
sustain the electromagnetic analogy, however, we need to require that
the gravitomagnetic charge be twice the gravitoelectric charge.  This
factor of 2 is a remnant of the spin-2 character of the original
field, while for a pure spin-1 field (i.e. the electromagnetic field)
the ratio of the magnetic charge to the electric charge is unity.

The equation of motion of a test particle of mass $m$ in this linear gravitational
field can be obtained from the variational principle $\delta \int {\cal L} dt=0$,
where ${\cal L}=-mcds/dt$ is given by

\begin{equation}
{\cal L} = - mc^{2} \left[ 1-{v^{2}\over c^{2}} - {2 \over c^{2}} \left( 1+
{v^{2}\over c^{2}} \right) \Phi + {4\over c^{3}} {\bf v} \cdot {\bf A}
\right]^{1/2},
\label{e12}
\end{equation}

\noindent using equation (\ref{e6}).  To linear order in $\Phi$ and ${\bf A}$, one can
write (\ref{e12}) as

\begin{equation}
{\cal L} = - mc^{2} \left( 1- {v^{2} \over c^{2}} \right)^{1/2} + m \gamma \left(
1+ {v^{2} \over c^{2}} \right) \Phi - {2 m \over c} \gamma {\bf v} \cdot
{\bf A}.
\label{e13}
\end{equation}

\noindent Let us note that the deviation of equation (\ref{e13}) from a free-particle
Lagrangian is given to lowest order in ${\bf v} / c$ by $m \Phi - 2m
{\bf A} \cdot {\bf v} /c$.  This deviation would be of the form
$j_{\mu}A^{\mu}$ in electrodynamics; therefore, the slow motion of the
test particle is very similar to that of a charged particle in
electrodynamics except that here $q_{E}=-m$ and $q_{B}=-2m$ as
expected.  It thus follows from the geodesic motion of a test particle
of mass $m$ far from the source in this gravitational background that
the canonical momentum of the particle is given approximately by ${\bf
  p}+(-2m/c){\bf A}$, where ${\bf p}$ is the kinetic momentum.  In
this electrodynamic analogy, the attractive nature of gravity is
reflected in our convention of positive gravitational charges for the
source and negative gravitational charges for the test particle.  The
gravitomagnetic charge is always twice the gravitoelectric charge as a
consequence of the tensorial character of the gravitational potentials
in general relativity.

The gauge transformations

\begin{equation}
\Phi \rightarrow \Phi - {1\over c} {\partial \psi \over \partial t}, \;\;\;\;\;
{\bf A} \rightarrow {\bf A} +2 \mbox{\boldmath $\nabla$} \psi,
\label{e14}
\end{equation}

\noindent leave the GEM fields (\ref{gem}) and 
hence the GEM equations (\ref{e1})-(\ref{e4})
invariant.  The Lorentz gauge condition (\ref{gauge}) is also satisfied
provided $\Box \psi = 0$.  However, the quantity $-q_{E} \Phi + q_{B}
{\bf A} \cdot {\bf v}/c$ in the Lagrangian is not invariant under the
gauge transformation (\ref{e14}).  The gauge invariance of this Lagrangian
is restored, however, if the gauge function $\psi$ is independent of
time, $\partial \psi/ \partial t = 0$.  In this case, we can start
from a coordinate transformation $t \rightarrow t - 4 \psi ({\bf
  x})/c^{3}$ in the metric (\ref{e6}) resulting in the gauge
transformations (\ref{e14}) with $\Phi$ left invariant.

The gravitational field corresponding to the metric (\ref{e6}) is
given by the Riemann curvature tensor

\begin{equation}
R_{\mu \nu \rho \sigma} = {1\over 2} (h_{\mu \sigma, \; \nu \rho} +
h_{\nu \rho, \; \mu \sigma} - h_{\nu \sigma,\; \mu \rho} - h_{\mu \rho, \nu
\sigma} ),
\label{e15}
\end{equation}

\noindent where $h_{00}=2\Phi /c^{2}$ and $h_{ij}=(2\Phi /c^{2}) \delta_{ij}$ are
gravitoelectric and of $O(c^{-2})$, while $h_{0i}=-2A_{i}/c^{2}$ is
gravitomagnetic and of $O(c^{-3})$.  The components of the curvature
tensor as measured by the {\it standard} geodesic observers are given
by $R_{\mu \nu \rho \sigma} \lambda^{\mu}_{(\alpha)}
\lambda^{\nu}_{(\beta)} \lambda^{\rho}_{(\gamma)}
\lambda^{\sigma}_{(\delta)}$, where $\lambda^{\mu}_{(\alpha)}$ is the
tetrad frame of the test observer.  In the linear approximation under
consideration here, $\lambda^{\mu}_{(\alpha)}$ is in effect equal to
$\delta^{\mu}_{\alpha}$ in the calculation of the measured curvature.
The components of this tensor may be expressed in the form of a
symmetric $6 \times 6$ matrix ${\cal R}=({\cal R}_{AB})$, where $A$
and $B$ range over (01, 02, 03, 23, 31, 12); hence,

\begin{equation}
{\cal R} = \begin{pmatrix} {\cal E} & {\cal B} \\ {\cal B}^{T} & 
{\cal S} \end{pmatrix}\,,
\label{e16}
\end{equation}

\noindent where ${\cal E}$ and ${\cal S}$ are symmetric $3 \times 3$ matrices and
${\cal B}$ is traceless.  We find that the electric and magnetic components of the
curvature are given by

\begin{equation}
{\cal E}_{ij}={1\over c^{2}} E_{j,i}+O(c^{-4}),
\label{e17}
\end{equation}

\begin{equation}
{\cal B}_{ij}=-{1\over c^{2}} B_{j,i} + {1\over c^{3}} \epsilon_{ijk} {\partial
E_{k} \over \partial t} + O(c^{-4}),
\label{e18}
\end{equation}

\noindent and the spatial components are given by

\begin{equation}
{\cal S}_{ij}=-{1\over c^{2}} E_{j,i} + {1 \over c^{2}} (\mbox{\boldmath $\nabla$}
\cdot {\bf E}) \delta_{ij} + O(c^{-4}).
\label{e19}
\end{equation}

That ${\cal B}$ is traceless is consistent with equation (\ref{e2}) and the
fact that ${\cal E}$ and ${\cal S}$ are symmetric is consistent with
equation (\ref{e3}) at $O(c^{-4})$.  It is therefore clear that
gravitoelectromagnetism permeates every aspect of general relativity:
the gravitational potentials (GEM potentials), the connection (GEM
field) and the curvature.  In the exterior of the rotating source, the
spacetime is Ricci-flat and hence ${\cal S} =-{\cal E}$, ${\cal E}$ is
traceless and ${\cal B}$ is symmetric.  These restrictions on the
curvature are consistent with the GEM field equations
(\ref{e1})-(\ref{e4}) in the source-free region.

The general treatment of gravitoelectromagnetism presented here has been based on
a certain approximate form of the linear gravitation theory and can be used in the
theoretical description of many interesting gravitational phenomena.  In
particular, we use this formalism below to investigate the microphysical
implications of the gravitomagnetic precession of spin.

\vspace{.15in}

\noindent {\it Free Fall is not Universal}

The assumption that all free test particles fall in the same way in a
gravitational field is reflected in general relativity via the
geodesic hypothesis.  That is, the worldline of a free test particle
is an intrinsic property of the spacetime manifold and is independent
of the intrinsic aspects of the particle.  In this way, general
relativity is a geometric theory of gravitation.  This circumstance
originates from the well-tested equality of inertial and gravitational
masses.

An important consequence of Einstein's geometric theory of gravitation
is the fact that an ideal test gyroscope would precess in the
gravitomagnetic field of a rotating source.  Here we pose the question
of whether all spins should ``precess'' like a gyroscope; evidently, the 
treatment of intrinsic spin would go beyond classical general relativity.  
It follows from the consideration of spin-rotation-gravity coupling that the
intrinsic spin of a particle (e.g. a nucleus) would couple to the
gravitomagnetic field of a rotating source (such as the Earth) via the
interaction Hamiltonian

\begin{equation}
{\cal H} = \mbox{\boldmath $\sigma$} \cdot \mbox{\boldmath $\Omega$}_{P}
\label{e20}
\end{equation}

\noindent such that the Heisenberg equations of motion for the spin would be
formally the same as that of an ideal test gyro \cite{6}.  
Intuitively, this interaction is due to the coupling of the 
gravitomagnetic dipole moment of the particle with the gravitomagnetic 
field just as would be expected from the electromagnetic analogy. It follows
from equation (\ref{e20}) that the particle is subject to a gravitational
Stern-Gerlach force given by

\begin{equation}
{\bf F} = - \mbox{\boldmath $\nabla$} (\mbox{\boldmath $\sigma$} \cdot
\mbox{\boldmath $\Omega$}_{P})
\label{e21}
\end{equation}

\noindent that is purely dependent upon its spin and not its mass and therefore
violates the universality of free fall.

The point is that a particle is in general endowed with mass and spin
in addition to other intrinsic properties; indeed, the irreducible
unitary representations of the inhomogeneous Lorentz group are
characterized by mass and spin.  In its interaction with a
gravitational field, the mass interacts primarily with the
gravitoelectric field while the spin interacts primarily with the
gravitomagnetic field.  Whereas the former dominant interaction is
consistent with the universality of free fall, the latter is not.  For
instance, the bending of light by the gravitational field of a
rotating source depends on the state of polarization of the radiation.
The differential deflection of polarized radiation by the Sun is too
small to be measurable at present.  The predicted violation is also
extremely small for a nucleus in a laboratory on the Earth: the weight
of the particle is $w=mg_{\oplus}(1\pm \epsilon)$, depending on whether the
spin is polarized vertically up or down and $\epsilon \sim 10^{-29}$.
Thus the predicted violation of the universality of free fall is
extremely small.

It may still be possible to measure this relativistic quantum gravitational effect
by detecting the change in the energy of a particle in the laboratory when its
spin is flipped.  This would require, for instance, significant refinements in
modern variations of NMR and optical pumping techniques, since

\begin{equation}
\hbar \Omega_{P} \sim \frac{\hbar GJ}{c^{2}R^{3}}={cJ \over R} \left(
\frac{L_{P}}{R} \right)^{2} \sim 10^{-28} {\rm eV}
\label{e22}
\end{equation}

\noindent is a factor of $10^{4}$ below the sensitivity of recent experiments
\cite{7}.  Here $L_{P}$ is the Planck length, $L_{P}^{2}=\hbar
G/c^{3}$, $J$ is the angular momentum of the Earth and $R$ is its
average radius.  The smallest detected energy shift is about
$10^{-24}$eV corresponding to a frequency shift of 2 nHz \cite{7}.  However,
it appears that significantly lower energy shifts may soon be
detectable \cite{8}.

To clarify the nature of the force (\ref{e21}), let us consider the motion
of a classical spinning test body in a {\it stationary} gravitational field.
Such a system is necessarily extended and thus couples to spacetime
curvature resulting in a Mathisson-Papapetrou force

\begin{equation}
F_{\alpha} = {c\over 2} R_{\alpha \beta \mu \nu} u^{\beta} S^{\mu \nu} = c
R_{\alpha \mu \beta \nu} u^{\beta} S^{\mu \nu},
\label{e23}
\end{equation}

\noindent where $S^{\mu \nu}$ is the spin tensor of the system, 
$u^\mu$ is the velocity vector such that $S^{\mu\nu}u_{\nu}=0$ and
the spin vector is given by

\begin{equation}
S_{\mu}={1\over 2} (-g)^{1/2} \epsilon_{\mu \nu \rho \sigma} u^{\nu} S^{\rho
\sigma}.
\label{e24}
\end{equation}

\noindent For the calculation of $F_{\alpha}$, it suffices to set, in the 
linear approximation, $u^{\alpha} \approx (1,\:0,\:0,\:0)$, $S^{0i}
\approx 0$ and $S^{ij} \approx - \epsilon^{ijk} S_{k}$. Then, 
$F_{0} \approx 0$ and

\begin{equation}
F_{i}\approx c {\cal B}_{ij} S^{j}=-{1\over c} B_{j,i} S^{j}
=- (\mbox{\boldmath $\Omega$}_{P})_{j,i} S^{j},
\label{e25}
\end{equation}

\noindent in agreement with equation (\ref{e21}).  Thus the existence of the force
(\ref{e21}) may be ascribed to the intrinsic nonlocality of a particle in
the quantum theory and hence the coupling of spin to the magnetic part
of the spacetime curvature in a stationary field.

It is important to remark here that our considerations are distinct
from proposals to measure the classical spin-spin force as discussed
in \cite{4}.  Our results ultimately follow from detailed
considerations of Dirac-type wave equations in the gravitational field
of a rotating mass (see the references cited in \cite{6}); however,
one can arrive at equations (\ref{e20})-(\ref{e21}) on the basis of certain
general arguments such as the local isotropy of space, the extended
hypothesis of locality and the gravitational Larmor theorem \cite{6}.

Assuming the approximate validity of equations (\ref{e20})-(\ref{e21}), it would be
difficult to imagine a basic gravitational theory founded purely on the
universality of free fall and Riemannian geometry.  However, such a theoretical
structure would clearly be an excellent effective theory in the macrophysical
domain.

\vspace{.15in}

\noindent {\it GEM Stress-Energy Tensor}

Let us imagine a congruence of geodesic test particles in a
gravitational field.  Taking one of the test particles as the
reference observer, how does the motion of the other neighboring test
particles appear to the fiducial observer?  The result is best
expressed in a Fermi coordinate system that is set up along the
reference worldline.  Let $X^{\mu} = (\tau, {\bf X})$ be the Fermi
coordinates of the test particles, while the reference observer is at
the origin of spatial Fermi coordinates.  Then the equation of motion
of the test particles is given by the generalized Jacobi equation

\begin{eqnarray}
\frac{d^{2}X^{i}}{d \tau^{2}}&+&^{F}\!R_{0i0j} X^{j}
+2\:^{F}\!R_{ikj0}V^{k}X^{j}+(2\:^{F}\!R_{0kj0}V^{i}V^{k} \nonumber \\
&+&{2\over 3}\: ^{F}\!R_{ikjl}V^{k}V^{l} + {2\over 3}\:
^{F}\!R_{0kjl}V^{i}V^{k}V^{l}) X^{j}=0,
\label{e26}
\end{eqnarray}

\noindent which is valid to first order in the relative separation ${\bf X}$ and
to all orders in the relative velocity ${\bf V}=d{\bf X}/d\tau$.  Here
$^{F}\!R_{\alpha \beta \gamma \delta}(\tau)$ are components of the curvature tensor
as measured by the fiducial observer, i.e. they are the projections of the Riemann
tensor onto the nonrotating tetrad frame of the reference observer.  Neglecting
the second and third order terms in the relative rate of separation, equation
(\ref{e26}) can be written as the GEM analog of the Lorentz force law

\begin{equation}
m \frac{d^{2}{\bf X}}{d\tau^{2}}=q_{E}{\bf E}+q_{B}{\bf V} \times {\bf B},
\label{e27}
\end{equation}

\noindent where $q_{E}=-m$, $q_{B}=-2m$ and

\begin{equation}
E_{i}=\:^{F}\!R_{0i0j}(\tau)X^{j}, \;\;B_{i}=- {1\over 2}
\epsilon_{ijk}\;^{F}\!R_{jk0l}(\tau) X^{l}.
\label{e28}
\end{equation}

It is important to notice that the spacetime interval in the
neighborhood of the reference worldline can be expressed in Fermi
coordinates as $-ds^{2}=\:^{F}\!g_{\mu \nu}\:dX^{\mu}dX^{\nu}$, where

\begin{equation}
^{F}\!g_{00}=-1-\:^{F}\!R_{0i0j}(\tau)X^{i}X^{j}+ \cdots ,
\label{e29}
\end{equation}

\begin{equation}
^{F}\!g_{0i}=-{2\over 3}\:^{F}\!R_{0jik}(\tau)X^{j}X^{k}+ \cdots ,
\label{e30}
\end{equation}

\begin{equation}
^{F}\!g_{ij}=\delta_{ij} - {1\over 3}\:^{F}\!R_{ikjl}(\tau)X^{k}X^{l} + \cdots .
\label{e31}
\end{equation}

\noindent Letting $^{F}\!g_{00}=-1+2 \Phi$ and $^{F}\!g_{0i}=-2A_{i}$, we find that

\begin{equation}
\Phi=-{1\over 2} \:^{F}\!R_{0i0j}X^{i}X^{j},\;\;\; A_{i}={1\over
3}\:^{F}\!R_{0jik}X^{j}X^{k},
\label{e32}
\end{equation}

\noindent so that the corresponding GEM fields using equation (\ref{gem}) agree with the
results in equation (\ref{e28}) to linear order in the separation ${\bf X}$.  One can
verify directly that $\mbox{\boldmath $\nabla$} \cdot {\bf B} = 0$ and
$\mbox{\boldmath $\nabla$} \times {\bf E} = 0$, so that the source-free pair of
Maxwell's equations are satisfied {\it along the reference worldline}.  Moreover, it
is possible to combine the GEM fields together to form a GEM Faraday tensor
$F_{\alpha \beta}$,

\begin{equation}
F_{\alpha \beta}=-\:^{F}\!R_{\alpha \beta 0l}X^{l},
\label{e33}
\end{equation}

\noindent such that $F_{0i}=-E_{i}$ and $F_{ij}=\epsilon_{ijk}B_{k}$ as in
standard electrodynamics.  Then the other pair of Maxwell's equations
is given by $F^{\alpha \beta},_{\beta}=4 \pi J^{\alpha}$, where
$J^{\alpha}(\tau, {\bf X})$ is easily obtained to linear order in
${\bf X}$ using equation (\ref{e33}).  $J^{\alpha}$ is a conserved current
such that $J_{\alpha}(\tau, {\bf 0})=-\:^{F}\!R_{\alpha 0}/4\pi$ along
the fiducial trajectory.  This treatment should be compared and
contrasted with the linear approximation developed above, in
particular, the GEM current is different here.

It is now possible to develop the classical field theory of the GEM field in the
Fermi frame; in particular, one can define the Maxwell stress-energy tensor and
the corresponding angular momentum for the GEM field.  Thus

\begin{equation}
{\cal T}^{\alpha \beta}(\tau,{\bf X})={1\over 4\pi} \left( F^{\alpha}\!{}_\gamma
F^{\beta \gamma}- {1\over 4} \eta^{\alpha \beta} F_{\gamma \delta}F^{\gamma \delta}
\right)
\label{e34}
\end{equation}

\noindent is the Maxwell stress-energy tensor for the GEM 
field that is quadratic in the spatial separation and vanishes at the
location of the fiducial observer.  Physical measurements do not occur
at a point, as already emphasized by Bohr and Rosenfeld \cite{9};
moreover, the fiducial observer is arbitrary here.  Therefore, a
physically more meaningful quantity is obtained by averaging equation
(\ref{e34}) over a sphere of radius $\epsilon L$ in the Fermi system.
Here $L$ is a constant invariant lengthscale associated with the
gravitational field.  We find that

\begin{equation}
<{\cal T}_{\alpha \beta}(\tau, {\bf X})>=\epsilon^{2}C_{0}L^{2} \tilde{T}_{\mu \nu
\rho \sigma} \lambda^{\mu}_{(\alpha)} \lambda^{\nu}_{(\beta)} \lambda^{\rho}_{(0)}
\lambda^{\sigma}_{(0)},
\label{e35}
\end{equation}
where $C_{0}$ is a constant numerical factor and

\begin{equation}
\tilde{T}_{\mu \nu \rho \sigma}={1\over 2} \left( R_{\mu \xi \rho
\zeta}R_{\nu \; \sigma}^{\; \: \xi \; \zeta} +R_{\mu \xi \sigma \zeta}
R_{\nu \; \rho}^{\; \: \xi \;\zeta} \right)  
-{1\over 4}g_{\mu \nu}R_{\alpha \beta \rho \gamma}R^{\alpha \beta \; \gamma}
_{\;\; \;\:\sigma}.
\label{e36}
\end{equation}

\noindent For a Ricci-flat spacetime, $\tilde{T}_{\mu \nu \rho \sigma}$ reduces to
the Bel-Robinson tensor $T_{\mu \nu \rho \sigma}$; in this case,
$R_{\alpha \beta \gamma \delta}$ reduces to the Weyl tensor $C_{\alpha
  \beta \gamma \delta}$ and in equation (\ref{e36}) $C_{\alpha \beta
  \rho \gamma}C^{\alpha \beta \; \gamma} _{\;\; \;\:\sigma} =
(K/4)g_{\rho \sigma}$ with $K=C_{\alpha \beta \gamma \delta}C^{\alpha
  \beta \gamma \delta}$.

The magnitude of $C_{0}$ depends on whether we average over the surface or the
volume of the sphere; in any case, one can always absorb $C_{0}$ into the
definition of $L$.  Thus the pseudo-local GEM stress-energy tensor may be defined
for any observer with the tetrad frame $\lambda^{\mu}_{(\alpha)}$ as

\begin{equation}
T_{(\alpha) (\beta)}=L^{2}\tilde{T}_{\mu \nu \rho \sigma} \lambda^{\mu}_{(\alpha)}
\lambda^{\nu}_{(\beta)} \lambda^{\rho}_{(0)} \lambda^{\sigma}_{(0)}.
\label{e37}
\end{equation}

\noindent In this way, the pseudo-local GEM energy density, Poynting flux and
stresses are defined up to a common multiplicative factor.

It is important to note that the spatial components of the curvature have been
ignored in our construction of the GEM tensor $T_{(\alpha)(\beta)}$.  For a
Ricci-flat spacetime, however, the spatial components of the curvature are simply
related to its electric components; therefore, the pseudo-local tensor defined via
equation (\ref{e37}) using the Bel-Robinson tensor contains the full (Weyl) curvature
tensor and is thus the {\it gravitational stress-energy tensor}.

It follows from a simple application of these results to the field of a rotating
mass that there exists a steady Poynting flux of gravitational energy in the
exterior field of a rotating mass.

\vspace{.15in}

\noindent {\it Oscillations of a Charged Rotating Black Hole}

Imagine a black hole of mass $M$, charge $Q$ and angular momentum $J$
that is perturbed by external radiation.  The black hole is stationary
and axisymmetric; therefore, the perturbation is expressible in terms
of eigenmodes ${\cal P} (r, \theta)\,{\rm exp}(-i\omega t$\\ $ +im_{j}
\phi)$, where ${\cal P}$ depends upon the frequency of the radiation,
the total angular momentum parameters of the eigenmode $(j, m_{j})$
and the spin of the external field.  It turns out that for a Fourier
sum of such eigenmodes, the response of the black hole far away and at
late times is dominated by a superposition of certain damped
oscillations of the form $A \,{\rm exp} (-i\omega t)$, where $\omega =
\omega_{BH}-i \Gamma_{BH}$ with $\Gamma_{BH} \geq 0$.  For these
quasinormal modes, the amplitude $A$ depends, among other things, on
the strength of the perturbation while $\omega$ depends only on the
black hole parameters $(M, Q, J)$.  Moreover, these black hole
oscillations are in general denumerably infinite and are numbered as
$n=0,1,2, \cdots$, such that $n=0$ is least damped and $\Gamma_{BH}$
increases with $n$.  The intrinsic ringing of a black hole is due to
the fact that once perturbed, the black hole undergoes characteristic
damped oscillations in order to return to a stationary state.

The fundamental modes of oscillations of black holes were originally
found by numerical experiments and initial attempts to explain the
numerical results via the properties of black hole effective
potentials were unsuccessful \cite{10}.  The solution of the problem
was first given around 1980 \cite{11}.  This work provided the
stimulus for many subsequent investigations by a number of authors
\cite{12}.  A detailed discussion of black hole oscillations is
contained in \cite{13}.

For the modes of oscillation of a charged rotating black hole, the
only reliable results are for $j\geq |m_{j}| \gg 1$.  Expressions for
$(\omega_{BH}, \Gamma_{BH})$ have been obtained in the case of
$j=|m_{j}| \gg 1$ for a general Kerr-Newman black hole; however, the
results have been generalized to the case of $j>|m_{j}| \gg 1$ only for a
slowly rotating charged black hole \cite{14}.  To express
$(\omega_{BH}, \Gamma_{BH})$ in terms of $(M,J,Q)$ in the latter case,
let $\omega_{\rm K} (r)=(Mr^{-3}-Q^{2}r^{-4})^{1/2}$ be the ``Keplerian''
frequency for the motion of a neutral particle in a circular geodesic
orbit of radius $r$ about a Reissner-Nordstr\"om black hole of mass
$M$ and charge $Q$.  Here we use Boyer-Lindquist type of coordinates
for the Kerr-Newman geometry.  Timelike circular geodesic orbits exist
down to a null orbit of radius $r_{N}$ such that
$2r_{N}=3M+(9M^{2}-8Q^{2})^{1/2}$.  Let
$\omega_{N}=\omega_{\rm K}(r_{N})$, then it can be shown that for a slowly
rotating black hole

\begin{equation}
\omega_{BH} \approx \pm j\omega_{N}+m_{j}\Omega_{N},
\label{e38}
\end{equation}

\noindent where

\begin{equation}
\Omega_{N}=\frac{J}{r_{N}^{3}} \left( 1- \frac{Q^{2}}{Mr_{N}} \right)
\frac{r_{N}+M}{r_{N}-M},
\label{e39}
\end{equation}

\noindent is an effective black hole rotation frequency.  The $(2j+1)$-fold
degeneracy in the spectrum of oscillations of the spherical black hole is removed
by its rotation.  We note that $\Omega_{N}$ is proportional to the gravitomagnetic
precession frequency at $r_{N}$.  Moreover,

\begin{equation}
\Gamma_{BH} \approx \left( n+ {1\over 2}\right) \left( 2-3 {M\over r_{N}}
\right)^{1/2} \left( \omega_{N} \mp {m_{j} \over j} Q_{*} \Omega_N \right),
\label{e40}
\end{equation}

\noindent where $n=0,1,2, \cdots$ is the mode number and $Q_{*}=6MQ^{2}/[r_N
(9M^{2}-8Q^{2})]$.  It is interesting to note that if the black hole is charged,
the rotation removes the degeneracy of the damping factor as well.  Moreover, in
the formulas (\ref{e38}) and (\ref{e40}) if $(\omega_{BH}, \Gamma_{BH})$ is a ringing mode,
then so is $(-\omega_{BH}, \Gamma_{BH})$.  These results are independent of the
spin of the perturbing field, since they are valid for states of high total
angular momentum $j \geq |m_{j}| \gg 1$.

\section{Structure of Time and Relativistic Precession}

Let us now return to the gravitomagnetic temporal structure around a 
rotating source. 
The gravitomagnetic clock effect involves a coupling between the {\em
orbital} motion of clocks and the rotation of the source.
On the other hand, the gravitomagnetic gyroscope precession occurs even for
a gyroscope at rest in the exterior field of a
rotating source.  Nevertheless, there is a general physical connection
between relativistic precession and temporal structure.
This is not surprising since the operational definition of time
ultimately involves counting a definite period and simple precession is uniform
periodic motion.
It is the purpose of this section to explain this relationship.  We do this
in several steps in the context of Kerr geometry
with
\begin{gather}
-ds^{2}=-dt^{2}+\frac
{\Sigma}{\Delta}\hspace{.05in}dr^{2}+\Sigma\hspace{.05in}
d\theta^{2}+(r^{2}+a^{2})\sin^{2}\theta\hspace{.05in} d\phi^{2}
\hspace{2cm}\nonumber \\
\hspace{7cm} +\frac{R_{g}r}{\Sigma}(c\hspace{.05in}dt-a\sin^{2}\theta
\hspace{.05in} d\phi)^{2}, 
\label{kerr}
\end{gather}
where $\Sigma=r^{2}+a^{2}\cos^{2}\theta$ and $\Delta=r^{2}-R_{g}r+a^{2}$.
Here $R_{g}=2GM/c^{2}$ is the gravitational radius of the
source and the Kerr parameter $a=J/Mc$ is a lengthscale characteristic of
the rotation of the source.  For $M=0$ and $a\neq0$,
the spacetime given by (\ref{kerr}) is flat as expected.  
For $a=0$ and $M\neq0$,
the metric (\ref{kerr}) represents the Schwarzschild
geometry.  Finally, for $a=0$ and $M=0$ we have the metric of an inertial
frame expressed in spherical coordinates
$(r, \theta, \phi)$.

Let us first imagine an accelerated observer in an inertial frame.  Suppose
that this observer carries along its worldline an
ideal pointlike test gyroscope so that there is no net torque on the
gyroscope and its spin axis is therefore nonrotating.  To
simplify matters, let us first assume that the path is a circle of radius
$r$ in the $(x,y)$-plane with its center at the origin
of coordinates.  According to the standard static observers in the inertial
frame, the accelerated observer moves with uniform
frequency $\omega_{*}\hat{{\bf z}}$.  The transformation between the
inertial frame and the rest frame of the rotating observer
involves a simple rotation of frequency $\omega_{*}$; therefore, from the
viewpoint of the standard (i.e. static) observers
in the rotating frame a natural operational way to keep the direction of the
gyroscope spin axis nonrotating is to imagine
fixing this axis at some initial time with respect to the axes of the {\em
rotating frame}, but then continuously rotating it
backward with frequency $\omega_{*}$.  In this way, the spin direction
would remain fixed in the inertial frame if the rotation
of the observer were virtual.  In reality, however, the
observer's proper time $\tau$ is related to  $t$ by
$d\tau=dt\hspace{.05in}(1-v^{2}/c^{2})^{1/2}$, where $v=r\omega_{*}$;
hence, the backward rotation of the spin occurs with respect to the rotating
observer's proper time, i.e. with frequency $\omega_{*}(dt/d\tau)$.  From
the standpoint of the standard inertial observers,
the time dilation causes the spin direction to overcompensate and hence the
spin direction is not fixed but precesses with the
Thomas precession frequency ${\bf \omega}_{{\rm T}}=-{\bf
\omega}_{*}(dt/d\tau)+{\bf \omega}_{*}$.  This amounts to a precession of
frequency
$\omega_{*}(\gamma-1)$ in a sense that is opposite to that of the rotation
of the comoving observer; moreover, the generalization
to arbitrary acceleration can be simply carried out by means of the Frenet
procedure.  That is, a Frenet frame can be set up
along the path of the observer in space; then, $\omega_{*}=v/R(t)$, where
$R(t)$ is the radius of the curvature at each
instant of time $t$.

The intimate connection between time dilation and Thomas precession in
Minkowski spacetime can be extended to a gravitational
field.  Therefore, let us imagine next that the motion described above is
the geodesic motion of a free test particle carrying
an ideal test gyroscope around a spherical mass $M$.  Let $\omega_{{\rm
K}}=d\phi/dt$ be the Keplerian frequency as perceived by
static inertial observers at infinity; then, $\omega^{2}_{{\rm
K}}=GM/r^{3}$, where $r$ is the Schwarzschild radius of the
circular orbit.  The proper frequency in this case is
$\omega=\Gamma\omega_{{\rm K}}$, where
$\Gamma=dt/d\tau=(1-3GM/c^{2}r)^{-1/2}$.
The gravitational time dilation involves the static ``gravitational
redshift'' effect of $-g_{00}=1-2GM/c^{2}r$ in the Schwarzschild geometry as well as
the azimuthal motion  $r^{2}(d\phi/dt)^{2}=GM/r$ resulting in the factor of
3 in $\Gamma$. This situation is reminiscent of the spin-orbit coupling in the
motion of the electron around the nucleus in the hydrogen atom; however, there
are subtle differences between the electromagnetic and gravitational
cases. In this case,
the spin precession frequency is given by the Fokker frequency
$\omega_{{\rm F}}=\omega-\omega_{{\rm K}}$, and the sense of
precession is in the \emph{same} sense as the orbital motion.  This
gravitational 
analog of the Thomas precession has a simple and
transparent explanation in terms of Einstein's principle of equivalence.
According to this heuristic principle, an observer
${\cal O}$ in a gravitational field is locally equivalent to an observer
${\cal O}^{\prime}$ in Minkowski spacetime with an acceleration
that is equal in magnitude but opposite in direction to the Newtonian
gravitational ``acceleration'' of ${\cal O}$.  The
gravitational (Fokker)  precession is thus locally equivalent to Thomas
precession  with the direction of acceleration reversed.
It follows that the Fokker precession is in the same sense as the orbital
motion.  For an arbitrary accelerated observer in a
gravitational field with velocity $u^{\mu}=dx^{\mu}/d\tau$ and acceleration
$a^{\mu}=Du^{\mu}/d\tau$, the nonrotating
equation of motion for the torque-free pointlike spin vector
$(u_{\mu}S^{\mu}=0)$ is
\begin{equation}
\frac{dS^{\mu}}{d\tau}+\Gamma^{\mu}_{\alpha\beta}\hspace{.05in}u^{\alpha}S^{\beta}=u^{\mu}a_{\nu}S^{\nu},
\label{spin}
\end{equation}
so that both Fokker and Thomas precessions would be present for accelerated
motion in Schwarzschild geometry.

The gravitomagnetic precession of a gyroscope is in a similar way related
to the temporal structure brought about by the
rotation of the source.  However, the situation here is more complicated
than the gravitoelectric Fokker precession since the
temporal structure is affected by the coupling of orbital motion with the
angular momentum of the source.

Specifically, let us imagine stable circular geodesic orbits in the
equatorial plane of the Kerr source (\ref{kerr}).  It can be shown
that
\begin{equation}
\frac{dt}{d\phi}=\frac{a}{c}\pm\frac{1}{2\pi}T_{{\rm K}} \hspace{.2in} ,
\hspace{.2in} \frac{d\tau}{d\phi}=\pm\frac{1}{2\pi}T_{{\rm K}}\hspace{.05in}
(1\pm2\alpha-3R_{g}/2r)^{1/2}\,,
\label{shown}
\end{equation}
where $T_{{\rm K}}=2\pi/\omega_{{\rm K}}$ is the Keplerian period of the
orbit and $\alpha=a\hspace{.05in}\omega_{{\rm K}}/c$.
Here the upper (lower) sign refers to a prograde (retrograde) orbit.  It
follows that the orbital period is given by
\begin{equation}
t_{\pm}=T_{{\rm K}}\hspace{.05in}(1\pm\alpha) \hspace{.2in}, \hspace{.2in}
\tau_{\pm}=T_{{\rm K}}\hspace{.05in}(1\pm2\alpha-3R_{g}/2r)^{1/2}\,.
\end{equation}
Let us first note that $t_{+}-t_{-}=4\pi a/c$ and
$\tau^{2}_{+}-\tau^{2}_{-}=4\alpha T^{2}_{{\rm K}}$.  Since
$\tau_{+}+\tau_{-}=2T_{{\rm K}}+O(c^{-2})$, we find that
$\tau_{+}-\tau_{-}\approx4\pi a/c$.  In fact,
$\tau_{+}-\tau_{-}$ monotonically decreases as a function of $r$ and
approaches $4\pi a/c$ as $r\rightarrow\infty$.  Thus
a prograde clock moves more slowly than a retrograde clock according to
comoving observers as well as the standard asymptotically
inertial observers at infinity; moreover, $\tau_{+}-\tau_{-}\approx4\pi
a/c$ for $r\gg R_{g}$.  This remarkable gravitomagnetic
clock effect has been discussed in some detail in recent publications
\cite{JMC}-\cite{WBB}.  This classical effect is in some sense the
gravitomagnetic analog of the topological Aharonov-Bohm effect; in fact,
let us note that far from a finite rotating source
$\tau_{+}-\tau_{-}$ is nearly a constant independent of the ``distance''
$r$ and the gravitational coupling constant $G$.
These aspects of the clock effect have been discussed in detail elsewhere
\cite{BFD}.

Let us now imagine two clocks moving in opposite directions on a stable
circular geodesic orbit of radius $r$ in the equatorial
plane of the Kerr source.  Suppose that at $t_{0}=0$, they are both at
$\phi_{0}=0$; let us denote the event at which the clocks
next meet again by $(t_{1},\phi_{1})$.  It follows from equation (\ref{shown})
that $2\pi t_{1}=\phi_{1}T_{{\rm K}}(1+\alpha)$ for
the prograde clock and $2\pi t_{1}=(2\pi -\phi_{1})T_{{\rm K}}(1-\alpha)$
for the retrograde clock.  Thus $\phi_{1}=\pi(1-\alpha)$
and $t_{1}=T_{{\rm K}}(1-\alpha^{2})/2$.  Moreover,
$\tau_{+}^{2}(\phi_{1})-\tau^{2}_{-}(\phi_{1})=\alpha T^{2}_{{\rm K}}
(\alpha^{2}+3R_{g}/2r)$, which is negligibly small for clocks in orbit
about astronomical sources in the solar system; in fact,
$\tau_{+}(\phi_{1})-\tau_{-}(\phi_{1})\sim O(c^{-4})$.  The next time the
clocks meet is at $(t_{2}, \phi_{2})$, which bears
the same relationship to $(t_{1},\phi_{1})$
as $(t_{1}, \phi_{1})$ to $(t_{0}, \phi_{0})$; therefore,
$\phi_{2}=2\pi(1-\alpha)$
and $t_{2}=T_{{\rm K}}(1-\alpha^{2})$.  In general, the {\em n}th time the
clocks meet is at $(t_{n}, \phi_{n})$ with
$\phi_{n}=n\pi(1-\alpha)$ modulo $2\pi$ and $t_{n}=nT_{{\rm
K}}(1-\alpha^{2})/2$.

Consider now the behavior of the diametrical line joining these events to
the origin of the spatial coordinates.
For $a=0$, i.e. in the Schwarzschild case, this line is fixed as the clocks
repeatedly meet at two diametrically opposite
points.  However, for $a\neq0$ the line precesses in the opposite sense as
the rotation of the source with the precession
frequency given approximately by (cf. Fig.\ 1)
\begin{equation}
\frac{\pi-\phi_{1}}{\tau_{+}(\phi_{1})}=\frac{GJ}{c^{2}r^{3}}+O(c^{-4}),
\end{equation}
which at this order is in agreement with the precession frequency of an
ideal torque-free gyroscope that is fixed in the
equatorial plane of the Kerr source \cite{mits70}.
\begin{figure}[htb]
\begin{center}
\includegraphics[width=.75\textwidth]{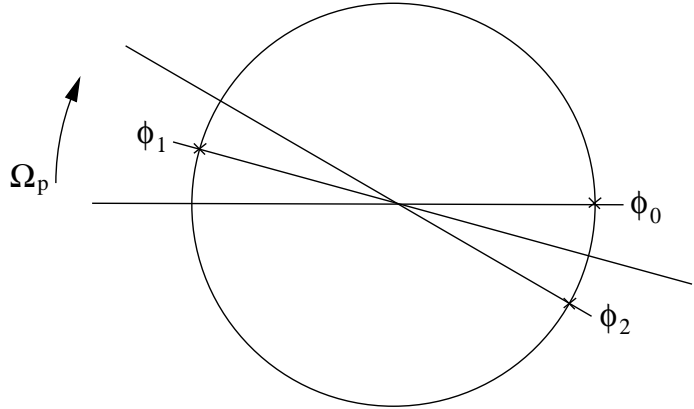}
\end{center}
\caption[]{Gravitomagnetic precession of the diametrical lines indicating the
points ($\phi_{0}, \phi_{1}, \phi_{2}, \ldots$) at which the clocks would 
meet.}
\label{eps1}
\end{figure}

Our treatment of the clock effect has been limited thus far to circular
orbits in the equatorial plane of the source.  Off this
plane, even circular orbits are not generally closed and the discussion of
the clock effect as well as its intimate connection
with the gravitomagnetic gyroscope precession becomes more complicated.  In
fact, the clock effect can be extended to such
orbits using the notion of {\em azimuthal closure} \cite{BFD}.

Finally, it should be mentioned that the general motion of an ideal
pointlike torque-free gyroscope in the Kerr field would, in
accordance with equation (\ref{spin}), involve Thomas and Fokker precessions as
well as a complicated gravitomagnetic motion that
consists of both precession and nutation.  Indeed, the notion of {\em
relativistic nutation} has been introduced in the post-Schwarzschild
approximation scheme in order to describe the nutational part of the
gravitomagnetic spin motion \cite{BDS}.  The complex
gravitomagnetic spin motion reduces to the simple (Schiff) precession in
the lowest post-Newtonian order.

\section{Clock Effect in the PPN Approximation}

In view of the possibility of detecting the gravitomagnetic clock effect 
via spaceborne clocks, it is interesting to develop the theory 
of the clock effect within the parametrized post-Newtonian 
(PPN) framework. The PPN formalism contains a set of parameters
that characterize different metric theories of gravitation in the
post-Newtonian approximation. 
The general form of the PPN metric  
is described in \cite{misn73,will93}; it includes theories
and effects that are not of primary interest for our treatment of the 
clock effect. Therefore we will start from a simplified PPN metric
of the form 
\begin{eqnarray}
g_{00} &=& -1+2U-2\beta U^2 \,,\label{metric1}\\
g_{0i}&=& -\frac{1}{4}(4\gamma+4+\alpha_1) H_i^* \,,\\
g_{ij} &=& (1+2\gamma U)\delta_{ij} \,,
\label{metric3}
\end{eqnarray}
which describes a rotating body in an underlying Cartesian coordinate system
$x^\mu=(ct,\boldsymbol{\varrho})$ with $\boldsymbol{\varrho}=(x,y,z)$.
In the following, we will use spherical coordinates $(\varrho, 
\theta, \phi)$; the isotropic radial coordinate $\varrho$ should not be 
confused with the Schwarzschild radial coordinate $r$. 
In general relativity, the PPN parameters $\alpha_1$, $\beta$ and 
$\gamma$ are given by $\alpha_1=0$ and $\beta=\gamma=1$.

The PPN metric (\ref{metric1})-(\ref{metric3}) is restricted to theories 
that exhibit conservation laws for total momentum and ignores the 
Whitehead and preferred-frame effects \cite{mash89}.
We assume that the gravitational source is a uniformly rotating
and nearly spherical body that is symmetric about the axis of 
rotation (i.e.\ the $z$-axis).
We are interested in the exterior gravitational field of the source
when its center of mass is at the origin of spatial coordinates. The
gravitoelectric potential $U(\varrho,\theta)$ is given in this 
case by \cite{5}
\begin{equation}
U(\varrho,\theta)=\frac{GM}{c^2\varrho}\Bigl[1+\sum_{n=2}^{\infty}
J_{n}\Bigl(\frac{\varrho_e}{\varrho}\Bigr)^{n} P_{n}(\cos\theta)\Bigr]\,,
\end{equation}
where $\varrho_e$ is the equatorial radius of the 
source, $M$ is in effect the asymptotically measured mass of the source, 
$P_n$ is the Legendre polynomial of degree $n$ and
\begin{equation}
J_n:=\frac{1}{M\varrho_e^n}\int \mu(\varrho', \theta') 
{\varrho '}^n P_n(\cos\theta')\, d^3\varrho' \,.
\end{equation}
Here $\mu$ denotes the effective mass-energy 
density of the source. In a 
similar way the multipole expansion of the gravitomagnetic 
vector potential $H_i^*$ can be expressed as \cite{5}
\begin{equation}
H_i^*(\varrho, \theta)=
\frac{G(\boldsymbol{J}\times\boldsymbol{\varrho})_i}{c^3 \varrho^3}
\Bigl[1+\sum_{n=1}^\infty K_{n} \Bigl(\frac{\varrho_e}{\varrho}\Bigr)^{n}
P'_{n+1}(\cos\theta)\Bigr]\,,
\end{equation}
where $\boldsymbol{J}=J \hat{{\bf z}}$ 
is in effect the asymptotically
measured angular momentum of the source and $P'_n(x)=dP_n(x)/dx$. Here
\begin{equation}
K_n:= \frac{2}{2n +3}\frac{M\varrho^2_e}{J}(L_n+J_{n+2})
\end{equation}
and
\begin{equation}
L_n:=\frac{1}{M\varrho_e^{n+2}}\int \mu(\varrho', \theta') 
{\varrho'}^{n+2}
P_n(\cos \theta') d^3\varrho' \,.
\end{equation}

The derivation of the clock effect involves the computation of the
proper time $\tau$ over a complete azimuthal cycle along geodesic
orbits about the source. For simplicity, we limit our discussion to
circular geodesic orbits in the equatorial plane, i.e.\ 
$\varrho= {\rm constant}$ and $\theta=\pi/2$. 

The radial geodesic equation, corresponding to a circular orbit in the 
equatorial plane, is given by 
\begin{equation}
\Gamma_{\alpha\beta}^{\varrho}\frac{dx^\alpha}{d\tau}\frac{dx^\beta}{d\tau}
=0\,,
\end{equation}
which can be written as 
\begin{equation}
\Bigr(\frac{c \,dt}{d\phi}\Bigl)^2+2\Bigl(\frac{c\, dt}{d\phi}\Bigr)
\frac{\Gamma_{0\phi}^\varrho}{\Gamma_{00}^\varrho} +
\frac{\Gamma_{\phi\phi}^\varrho}{\Gamma_{00}^\varrho} =0\,.
\label{can}
\end{equation}
It is straightforward to show that $\Gamma^\varrho_{0\phi}/\Gamma^\varrho_{00}
=g_{0\phi,\varrho}/g_{00,\varrho}$ and $\Gamma^\varrho_{\phi\phi}/
\Gamma^\varrho_{00} =g_{\phi\phi ,\varrho}/g_{00,\varrho}$. Using equations
(\ref{metric1})-(\ref{metric3}), we find that
\begin{equation}
g_{0\phi}=-\frac{1}{4}\Bigl(4\gamma+4+\alpha_1\Bigr) H(\varrho, \theta)
\sin^2 \theta \,,
\end{equation}
where
\begin{equation}
H(\varrho, \theta)=
\frac{GJ}{c^3 \varrho}
\Bigl[1+\sum_{n=1}^\infty K_{n} \Bigl(\frac{\varrho_e}{\varrho}\Bigr)^{n}
P'_{n+1}(\cos\theta)\Bigr]\,,
\end{equation}
and $g_{\phi\phi}=\varrho^2(1+2\gamma U)\sin^2\theta$. The solution 
of equation (\ref{can})
can then be written as 
\begin{eqnarray}
\frac{dt}{d\phi}&=&  \pm \Bigl| \frac{c^2}{\varrho} 
U_{,\varrho}\Bigr|^{-\frac{1}{2}}\Bigr[1+(\beta+\gamma)
U+\frac{1}{2}\gamma\varrho U_{,\varrho}\Bigr] \nonumber \\
&& +\frac{1}{8c}\Bigl(4\gamma+4+\alpha_1\Bigr)\frac{H_{,\varrho}}{U_{,\varrho}}
+ O({c^{-3}})\,.
\label{preli}
\end{eqnarray}
It follows from the PPN metric $-c^2 d\tau^2=c^2g_{00} dt^2+2cg_{0\phi}
dt d\phi+g_{\phi\phi}d\phi^2$ that 
\begin{equation}
\Bigl(\frac{d\tau}{d\phi}\Bigr)^2=(1-2U)\Bigl(\frac{d t}{d\phi}\Bigr)^2
-\frac{1}{c^2}\varrho^2
+ {O}({c^{-3}})\,.
\end{equation}
Using equation (\ref{preli}), we find after some algebra that
\begin{eqnarray}
\frac{d\tau}{d\phi}& =& \pm \Bigl| \frac{c^2}{\varrho} 
U_{,\varrho}\Bigr|^{-\frac{1}{2}}\Bigl[1+(\beta+\gamma-1)U +
\frac{1}{2}\varrho(\gamma U_{,\varrho} -|U_{,\varrho}|)\Bigr] \nonumber \\
&& +\frac{1}{8c}\Bigl(4\gamma+4+\alpha_1\Bigr) 
\frac{H_{,\varrho}}{U_{,\varrho}}+ O({c^{-3}})\,.
\end{eqnarray}
Integration of this equation immediately yields $\tau_{\pm}$; hence,
\begin{equation}
\tau_+-\tau_-=\frac{\pi}{2c}\Bigl(4\gamma+4+\alpha_1\Bigr) 
\frac{H_{,\varrho}}{U_{,\varrho}}+ O({c^{-3}})
\end{equation}
gives the gravitomagnetic clock effect within the restricted PPN framework
adopted here. The explicit dependence of the gravitomagnetic 
clock effect on the PPN parameters is through the proportionality factor of
$(4\gamma+4+\alpha_1)$; in fact, the clock effect has this feature 
in common with other main gravitomagnetic effects \cite{mash89}.

It is interesting to note that in general relativity the gravitomagnetic 
clock effect in the post-Newtonian approximation is given by 
\begin{equation}
\tau_+-\tau_-\approx 4\pi\frac{J}{Mc^2}\Bigl[1+\Bigl(\frac{3}{2}J_2-
\frac{9}{2} K_2\Bigr)\frac{\varrho_e^2}{\varrho^2}\Bigr]\,,
\end{equation}
when the source is assumed to be
symmetric about the equatorial plane and all moments higher than the
quadrupole are neglected. Using data given in
\cite{5}, we find that for the Earth $J_2\approx -10^{-3}$ and 
$K_2\approx -10^{-3}$, so that $(3/2)J_2-(9/2) K_2\approx 3\times 10^{-3}$
gives the relative contribution of the oblateness of the Earth to the 
clock effect for a near-Earth equatorial orbit.

\section{Detection of the Gravitomagnetic Temporal Structure}

According to Eq.\ (\ref{shown}) and the discussion following it, the orbital
motion of free clocks around a rotating mass gives rise to the
gravitomagnetic clock effect which shows up in the difference between
the proper orbital periods of co- and counter-orbiting clocks. This is
given by $4\pi a/c$ for equatorial trajectories. Inserting the
specific angular momentum of the Earth ($a\sim 3$ m) yields an
amazingly "large" value of $\tau_+-\tau_- \sim 10^{-7}$s.

Despite this seemingly large effect, the actual measurement of this
time difference encounters severe practical difficulties. Since the
two clocks are assumed to move along opposite but identical orbits,
their Kepler periods exactly cancel upon forming the difference
$\tau_+-\tau_-$, thereby revealing the gravitomagnetic clock
effect. In reality, however, clocks cannot be injected into
identical trajectories and the resulting difference in the Kepler
periods will readily exceed the time difference induced by the
rotation of the Earth. Since for near-Earth orbits a radial 
separation of 0.1 mm of the
clocks involves a time difference in the Kepler periods of the same
order of magnitude as the gravitomagnetic clock effect, the position
of the clocks has to be known at the submillimeter level in order to
filter the effect which is caused by the rotation of the Earth out of
the data. Similary, as the satellite moves just under 1 mm within 
$10^{-7}$s along its track (or $\sim 10^{-2}$ milliarcseconds in
angular distance), the azimuthal position has to be known at the
same accuracy as the radial one.  
On the other hand, the gravitomagnetic time difference
accumulates with the number of revolutions and after hundreds or
thousands of periods a knowledge of the position of the clocks at the
centimeter level will be sufficient to overcome the difference in the
Kepler periods.

Another difficulty arises from the determination of all the forces
that act on the satellites carrying the clocks.  Since the period of
revolution for orbits of $\sim 10^3$ km altitude is of the order of
$\sim 10^4$ s, accelerations as weak as $10^{-12}$ m/s$^2$ will
already cover the gravitomagnetic clock effect. Among these forces,
gravitational perturbations due to the nonsphericity of the Earth,
solid and ocean Earth tides as well as the interaction with the Sun,
Moon and planets will cause the most significant deviations from an
ideal orbit. Depending on the altitude of the satellites, the
atmospheric drag effect can also considerably change the shape of the
orbit.  Moreover, this latter effect is quite difficult to model because it
strongly depends on the atmospheric density which is not only
correlated to the orbital height but also subject to temporal
variations. Other non-gravitational perturbations like solar and
terrestrial radiation pressure, thermal thrust, charged particle drag
etc. must also be taken into account despite their less distinct
influence, since they likely induce accelerations in excess of
$10^{-12}$ m/s$^2$.

In practice, the effect of all these perturbations will be modeled by
determining a precise orbit based on the actual spacecraft
observations. This will be accomplished by generating an orbital
trajectory following Newton's equations of motion and by including all
perturbing forces acting on the satellite, using the most accurate
models available. In the next step, this predicted orbit has to be
best fitted to the one observed, where some force parameters may be
solved for during the orbit adjustment process in order to obtain an
improved or tailored force model for the specific mission. From the
resulting precise orbit the effects of the individual perturbations
are removed step by step thus yielding a quasi-Keplerian orbit,
but still carrying the signatures of the relativistic effects.
Finally, a comparison with the corresponding clock predictions for an
appropriate synthetic orbit is performed which is expected to confirm
the clock effect being investigated.

Therefore, in order to meet the very stringent conditions for the
observation of the clock effect, many tiny perturbing sources have to
be considered and investigated that are usually absent in most of
the present orbit determination systems.

\section{Quantum Origin of Inertia}
The sign of the gravitomagnetic clock effect has a remarkable consequence
that will be elucidated in this section.  It follows
from $t_{+}>t_{-}$ and $\tau_{+}>\tau_{-}$ that the uniform motion of the
prograde clock is slower than that of the retrograde
clock.  Thus in comparison with motion around a nonrotating mass, a
rotating mass would ``drag'' free test particles along
such that it would take longer (shorter) to go once around it on an
equatorial circular orbit in the prograde (retrograde)
direction.  We may call this circumstance virtual ``inertial antidragging,'' 
since it is the exact opposite of what would be expected
on the basis of the so-called ``inertial dragging'' \cite{AE1}.  In fact, as
Fig.\ 2 clearly demonstrates, for a given $r$ (i.e.
fixed orbital radius), the faster the Kerr source spins, the slower the
prograde motion and the faster the retrograde motion.

\begin{figure}[bth]
\begin{center}
\includegraphics[width=.7\textwidth]{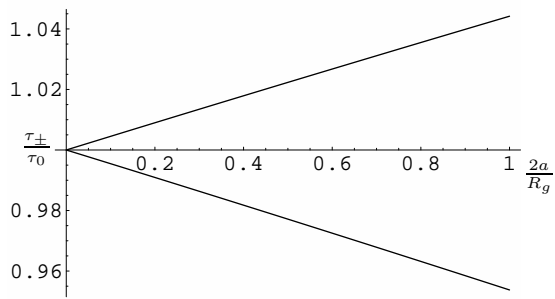}
\put(-245,90){$\frac{\tau_{\pm}}{\tau_0}$} 
\put(-55,80){$\frac{2a}{R_g}$}
\end{center}
\vspace{-1cm}
\caption{Plot of the clock rate versus the rotation of the source.  Imagine
an ensemble of Kerr fields with the same mass but
different angular momenta.  For a fixed stable circular geodesic orbit with
``radius'' $r$, $\tau_{+}(\tau_{-})$ monotonically
increases (decreases) over the ensemble with increasing angular momentum of
the source.  Stable orbits of this type occur for
$r\geq r_{\pm}$, where $r_{\pm}-3R_{g}\pm
4a\hspace{.05in}(2R_{g}/r_{\pm})^{1/2}=3a^{2}/r_{\pm}$.  Let us note that
$r_{-}\geq r_{+}$ and the equality
occurs for $a=0$ and $r_{\pm}=3R_{g}$; moreover, $r_{\pm}/a\rightarrow
(3)^{1/2}$ as $2a/R_{g}\rightarrow\infty$.  The graph illustrates
the behavior of $\tau_{\pm}$ for an ensemble of Kerr black holes with
$2a\leq R_{g}$; for $2a=R_{g}$, $r_{+}=a$ and $r_{-}=9a$.  In the
graph, $\tau_{0}=\tau_{\pm}(a=0)$ and the radius $r$ is chosen to be $5R_{g}$.}
\label{eps2}
\end{figure}

Rotational or translational inertial dragging refers to the circumstance
that an accelerating mass would somehow induce acceleration in
the same sense in nearby masses as a consequence of the so-called ``Mach's
principle.''  The clock effect indicates that precisely
the opposite situation is predicted for rotational motion in the
equatorial plane by the general
theory of relativity.  Translational inertial dragging
has been discussed by a number of authors \cite{OGr}; again, such notions
are foreign to the standard geometric interpretation of general
relativity \cite{BM2}.  In general relativity, accelerated motion is
absolute in the sense that it is nonrelative. Thus the term
``absolute'' as employed here only signifies the 
opposite of the term ``relative'' and is devoid of any metaphysical
connotations. The gravitomagnetic
clock effect and the gyroscope precession indicate the absolute rotation
of the source.  That is, a direct gravitomagnetic
verification of Einstein's theory of gravitation---e.g. via NASA's
GP-B---would constitute observational proof that the rotation
of the Earth is absolute and not merely relative to the distant matter of
the universe \cite{BM3}.

Mach's profound analysis of the foundations of Newtonian mechanics
occasioned a thorough re-examination of the basic classical
notions of space, time and motion that had been prevalent since Newton
provided a rational basis for the Copernican revolution
and Kepler's laws of planetary motion.  This re-evaluation culminated in
Einstein's theory of {\em general relativity}.  It is
therefore of great importance to recognize that general relativity---which
agrees with all experimental data to date---does not contain
the idea of relativity of arbitrary motion.  That is, this concept --- which was so crucial in the historical
development of Einstein's theory --- is absent in the standard geometric
interpretation of general relativity in the sense that it is neither a part of
the foundations of the theory nor follows from it. The re-emergence of absolute
motion may be taken to mean that general relativity
is still not completely devoid of certain ``metaphysical'' elements.
Should general relativity therefore be modified or abandoned
in favor of a theory based on the relativity of arbitrary motion?  To do so
would be unwise.  One should recognize instead that
physics has progressed far beyond the early days of relativity theory and
the observational successes of general relativity must
now be integrated within a quantum framework that involves the vacuum state
of microphysics as well as the rest frame of the cosmic
microwave background radiation.

Mach noted that in Newtonian mechanics, the intrinsic state of a classical
particle characterized by its mass $m$ has no direct
connection with the extrinsic state of the particle characterized by its
position and velocity (${\bf x,v}$) in absolute space
and time.  Hence the same extrinsic state can be occupied by other masses
comoving with the particle.  Thus an observer can change its
perspective by comoving with each particle in turn. In Newtonian mechanics,
the particles are thus ``placed'' on the absolute space and time continuum
and remain external to it.  On the other hand, 
classical particles are ``connected'' to each other via interactions such
as gravity and electromagnetism.  Mach therefore
concluded that only the motion of a particle relative to other
particles should have ultimate physical significance.
Mach's basic analysis has been restated in modern form in \cite{HLi}.

In classical physics, motion takes place via classical particles as well as
electromagnetic waves.  It appears that Mach did
not extend his analysis of classical particle motion to electromagnetic
wave propagation; in this connection, the issues that
arise in the examination of the historical record are briefly mentioned in
the Appendix.  Let us therefore apply Mach's argument
to the motion of electromagnetic waves.  The intrinsic aspects of the wave
are its amplitude, period, wavelength and polarization, which
therefore characterize its intrinsic state.
The extrinsic state of the wave is given by its wave function $\Psi(t,{\bf
x})$ in absolute time and space, and we note that the
wave's intrinsic state is directly related to its extrinsic state, i.e.
electric and magnetic field components, since the former cannot be defined
independently of the
latter.  The conclusion is that the motion of classical electromagnetic
waves is absolute, i.e. nonrelative.

Classical motion can be either relative or absolute.  In Einstein's
discussion of the so-called ``Mach's principle,'' only
``ponderable'' masses are considered \cite{AE1}, whereas classical motion
occurs via classical particles as well as electromagnetic waves.
It is natural to think of the motion of classical particles (i.e.
``ponderable'' masses) as relative, since one can change
one's perspective by moving with each mass in turn.  In the same sense, the
motion of electromagnetic waves must be considered
absolute due to its observer-independent status.  The development of
these simple notions taking due account of 
wave-particle duality leads to the principle of
complementarity of absolute and relative motion \cite{BM4}. 
In this connection, let us note that Mach's analysis of classical
particle motion may be restated in terms of the complete kinematic independence
of the absolute position $\bf {x}$ of a particle from its momentum
${\bf p}=m{\bf v}$.  In contrast, quantum kinematics can be consistently
formulated only by imposing the fundamental quantum condition on the operators
characterizing the position and momentum of a particle in absolute space and
time, i.e. $[\hat{x}_{j}, \hat{p}_{k}]=i\hbar \delta_{jk}$.  For instance, in the
nonrelativistic motion of a free particle in the Heisenberg picture
${\bf \hat{p}} = m{\bf \hat{v}}$ and $[\hat{x}_{j}, \hat{v}_{k}]=i\hbar
m^{-1}\delta_{jk}$.  Thus in contrast to the situation in classical mechanics,
the mass of a particle is related to its position and velocity in quantum
mechanics due to the fact that the particle has wave characteristics as well. 
This idea naturally extends to the specific orbital angular momentum of the
particle, $\hat{l}_{i}=\epsilon_{ijk}\hat{x}_{j}\hat{v}_{k}$, so that $[\hat{l}_{j},
\hat{l}_{k}] = i\hbar m^{-1}\epsilon_{jkn}\hat{l}_{n}$.  In the limit of an infinitely massive
particle, the connection disappears and the position and velocity commute; that is, one recovers
classical mechanics when the system is so massive that the perturbation due to an act of
observation on the system is negligible and the system therefore behaves classically.

Mach's argument involves classical quantities (c-numbers), whereas the quantum
condition involves operators (q-numbers); nevertheless, the quantum condition
implies that the intrinsic and extrinsic aspects of the particle are directly
related through Planck's constant.  For instance, in the Schr\"{o}dinger picture
the extrinsic state of the particle is given by the wave function $\Psi(t, {\bf x})$ and the
Schr\"{o}dinger equation for $\Psi$ involves $m$, which characterizes the intrinsic
state of the particle in Mach's analysis.  The relationship under discussion
here is not merely formal but can be verified observationally.  In fact, this
kinematic connection is particularly well illustrated by the example of a free
particle passing through a slit.  The resulting diffraction angle is inversely
proportional to the mass of the particle, so that the diffraction is absent in
the limit of large mass and the particle behaves classically.  To the extent
that classical mechanics can be thought of as a limiting form of quantum
mechanics, the epistemological problem of Newtonian mechanics --- so clearly
brought out by Mach --- disappears.  That is, the problem of the origin of
inertia is resolved through the wave nature of matter.  

Thus far the inertial mass of the particle has provided the quantum connection
to the inertial reference frames of Newtonian mechanics.  The invariance group
of Minkowski spacetime is the Poincar\'{e} group whose irreducible unitary
representations can be described in terms of mass and spin.  Thus in the
relativistic theory the inertial properties of a particle are characterized by
mass and spin.  The inertial properties of intrinsic spin have been discussed
in \cite{mash1}.  

Inertia has its origin in the fact that matter is intrinsically extended in
space and time and through this nonlocality inertial reference frames can be
``recognized''; then, a physical system tends to preserve its state with
respect to such frames.  This is beautifully illustrated by experiments involving
macroscopic quantum systems that have phase coherence, such as the recent
demonstration of Earth's absolute rotation via superfluid He$^4$
\cite{schwab}.
The quantum aspects of the 
origin of inertia are further developed in \cite{BM5}.

\section{Discussion}

In this paper we have examined some of the main theoretical aspects of
gravitomagnetism in general relativity.  The influence of the proper rotation
of a  source on the spacetime structure can be studied in various ways. 
Attention has been focused here on certain features of the gravitomagnetic
clock effect and its relation with the gravitomagnetic gyro precession. 
However, other approaches exist and should be mentioned.  The detection of the
gravitomagnetic field of the Earth via the Lense-Thirring precession of
satellite orbits has been investigated by Ciufolini {\it et al.}
\cite{ciufolini}. 
Moreover, gravitomagnetic effects in the spacetime curvature can be measured in
principle using gravity gradiometry \cite{theiss}.  In this connection, it is
interesting to note that gravity gradiometers of high sensitivity 
that are based on atomic interferometry are being
developed for space applications \cite{snadden,borde}.

\section*{Appendix: Mach and the Absolute Motion of Light}

Newton's introduction of the concepts of absolute space and time was truly
revolutionary in his day and allowed him to formulate
the basic classical laws of particle motion.  Leibniz \cite{GLe} and
Berkeley \cite{GBe} criticized the notions of absolute space and absolute
time
and emphasized instead the idea of relativity of all motion.  Later,
however, Maxwell \cite{JCM} relied
on absolute space and time for his fundamental extension of the Newtonian
ideas of motion to electromagnetic field propagation.
On the other hand, Mach revived the principle of relativity of all motion
on the basis of a profound analysis of the foundations
of classical mechanics \cite{EM1}.  Mach's work played a significant role
in Einstein's development of the theory of relativity \cite{AE2}.

Mach's deep physical treatment of the relativity of classical particle
motion was motivated by his epistemological stance on the relativity
of all measurement.  According to Mach, the result of a measurement is
the establishment of a {\em relation} and not of "absolute" notions,
since in Mach's view the latter refer to processes or objects that
are not
empirically verifiable in principle \cite{EM2}.  Mach's analysis of the
relativity of particle motion in his great work on classical mechanics
\cite{EM1} was not extended to electromagnetic wave motion in his
later work on physical optics \cite{EM3}.  In this book, Mach
discussed the wave theory of light as well as the speed of light;
however, he apparently made no attempt to put these in the context of
his epistemological stance on the relativity of all motion.  There is
no evidence that Mach ever wavered in his opposition to absolute
motion \cite{JTB}.  However, a number of Mach's contemporaries pointed
out the absolute character of the constancy of the speed of light and
were troubled by the fact that this aspect of the relativity theory
was in conflict with the relativity of all motion.  Among the
physicists and philosophers who raised such doubts about the
epistemological stance of {\em the theory of relativity} one can
mention Friedrich Adler, Hugo Dingler, Philipp Frank, Anton Lampa and
Joseph Petzoldt. Although it is claimed in the book of Blackmore \cite{JTB} 
that Mach rejected the principle of the constancy of 
the speed of light because it was in contradiction to his 
phenomenalistic epistemology due to its constant validity independent 
of all sensations and conscious data, there is actually no evidence 
that Mach ever directly or indirectly 
commented on the constancy of the speed of light
\cite{GWo}. A historical analysis of the 
situation and the influence of these criticisms on Mach can be found 
in the monographs of Blackmore \cite{JTB} and Wolters 
\cite{GWo}.

An exposition of the reasons for the supposed opposition to the
theory of relativity based on epistemological considerations and
experimental facts was promised to appear in a sequel to Mach's book
on optics \cite{EM3} in collaboration with his son Ludwig, but this was
never published. Although the preface to the
"Optics" is generally regarded as the most obvious evidence of Mach's
reluctance to accept the relativity theory, there is every reason to believe that
it was written by Ludwig only after the death of his father and
expresses Ludwig's opinion on the theory of relativity, despite the fact that Ernst
Mach is stated to be the author of this preface. More on this conjecture can be found in
\cite{GWo}. 

Finally, the position of Mach vis-\`{a}-vis the theory of relativity
is also discussed in the paper of Thiele \cite{JTh}.

%

\end{document}